\documentclass[prb,twocolumn,floatfix,preprintnumbers,amsmath,amssymb,superscriptaddress,showpacs]{revtex4-1}
\usepackage{amsmath}

\usepackage{graphicx}
\usepackage{graphics}
\usepackage{dcolumn}
\usepackage{bm}
\usepackage{xcolor}

\begin{document}

\title{Superconductivity suppression of Ba$_{0.5}$K$_{0.5}$Fe$_{2-2x}$$M$$_{2x}$As$_2$ single crystals by substitution of transition-metal ($M$ = Mn, Ru, Co, Ni, Cu, and Zn)}

\author{J.~Li}
\email{Li.Jun@nims.go.jp}
 \affiliation{Superconducting Properties Unit, National Institute for Materials Science, 1-1 Namiki, Tsukuba, Ibaraki 305-0044, Japan.}
 \affiliation{Department of Chemistry, Graduate School of Science, Hokkaido University, Sapporo, Hokkaido 060-0810, Japan.}

 \author{Y. F.~Guo}
 \affiliation{Superconducting Properties Unit, National Institute for Materials Science, 1-1 Namiki, Tsukuba, Ibaraki 305-0044, Japan.}
 \affiliation{International Center for Materials Nanoarchitectonics (MANA), National Institute for Materials Science, 1-1 Namiki, Tsukuba, Ibaraki 305-0044, Japan.}
 \author{S. B.~Zhang}
 \affiliation{Superconducting Properties Unit, National Institute for Materials Science, 1-1 Namiki, Tsukuba, Ibaraki 305-0044, Japan.}
 \affiliation{International Center for Materials Nanoarchitectonics (MANA), National Institute for Materials Science, 1-1 Namiki, Tsukuba, Ibaraki 305-0044, Japan.}
 \author{J.~Yuan}
 \affiliation{Superconducting Properties Unit, National Institute for Materials Science, 1-1 Namiki, Tsukuba, Ibaraki 305-0044, Japan.}
 \author{Y.~Tsujimoto}
 \affiliation{International Center for Materials Nanoarchitectonics (MANA), National Institute for Materials Science, 1-1 Namiki, Tsukuba, Ibaraki 305-0044, Japan.}
 \author{X.~Wang}
 \affiliation{Superconducting Properties Unit, National Institute for Materials Science, 1-1 Namiki, Tsukuba, Ibaraki 305-0044, Japan.}
 \affiliation{Department of Chemistry, Graduate School of Science, Hokkaido University, Sapporo, Hokkaido 060-0810, Japan.}
 \author{C. I.~Sathish}
 \affiliation{Superconducting Properties Unit, National Institute for Materials Science, 1-1 Namiki, Tsukuba, Ibaraki 305-0044, Japan.}
 \affiliation{Department of Chemistry, Graduate School of Science, Hokkaido University, Sapporo, Hokkaido 060-0810, Japan.}
 \author{Y.~Sun}
 \affiliation{International Center for Materials Nanoarchitectonics (MANA), National Institute for Materials Science, 1-1 Namiki, Tsukuba, Ibaraki 305-0044, Japan.}

 \author{S.~Yu}
 \affiliation{Superconducting Properties Unit, National Institute for Materials Science, 1-1 Namiki, Tsukuba, Ibaraki 305-0044, Japan.}

 \author{W.~Yi}
  \affiliation{International Center for Materials Nanoarchitectonics
(MANA), National Institute for Materials Science, 1-1 Namiki,
Tsukuba, Ibaraki 305-0044, Japan.}

 \author{K.~Yamaura}

  \affiliation{Superconducting Properties Unit, National Institute for Materials Science, 1-1 Namiki, Tsukuba, Ibaraki 305-0044, Japan.}
   \affiliation{Department of Chemistry, Graduate School of Science, Hokkaido University, Sapporo, Hokkaido 060-0810, Japan.}
 \affiliation{JST, Transformative Research-Project on Iron Pnictides (TRIP), 1-1 Namiki, Tsukuba, Ibaraki 305-0044, Japan.}
 \author{E.~Takayama-Muromachiu}
  \affiliation{Department of Chemistry, Graduate School of Science, Hokkaido University, Sapporo, Hokkaido 060-0810, Japan.}
   \affiliation{International Center for Materials Nanoarchitectonics (MANA), National Institute for Materials Science, 1-1 Namiki, Tsukuba, Ibaraki 305-0044, Japan.}
 \affiliation{JST, Transformative Research-Project on Iron Pnictides (TRIP), 1-1 Namiki, Tsukuba, Ibaraki 305-0044, Japan.}
 \author{Y.~Shirako}
 \affiliation{Department of Chemistry, Gakushuin University, 1-5-1 Mejiro,
Toshima-ku, Tokyo 171-8588, Japan.}
 \author{M.~Akaogi}
 \affiliation{Department of Chemistry, Gakushuin University, 1-5-1 Mejiro,
Toshima-ku, Tokyo 171-8588, Japan.}
 \author{H.~Kontani}
 \affiliation{Department of Physics, Nagoya University, Furo-cho, Nagoya 464-8602, Japan.}
\date{\today}

\begin{abstract}
We investigated the doping effects of magnetic and nonmagnetic
impurities on the single-crystalline $p$-type
Ba$_{0.5}$K$_{0.5}$Fe$_{2-2x}$$M$$_{2x}$As$_2$ ($M$ = Mn, Ru, Co,
Ni, Cu and Zn) superconductors. The superconductivity indicates
robustly against impurity of Ru, while weakly against the impurities
of Mn, Co, Ni, Cu, and Zn. However, the present $T$$_c$ suppression
rate of both magnetic and and nonmagnetic impurities remains much
lower than what was expected for the s$_{\pm}$-wave model.  The
temperature dependence of resistivity data is observed an obvious
low-$T$ upturn for the crystals doped with high-level impurity,
which is due to the occurrence of localization. Thus, the relatively
weak $T$$_c$ suppression effect from Mn, Co, Ni, Cu, and Zn are
considered as a result of localization rather than pair-breaking
effect in $s$$_{\pm}$-wave model.

\end{abstract}


\maketitle
\section[center]{introduction}
The existence of Fe-based superconductor family arouses unexpected
rapidly development\cite{1,2,3}, for that is not only a second class
of high-$T$$_c$ superconductors after the cuprate superconductors,
but also because it is highly promising to understand the
superconductivity (SC) mechanism of high-$T$$_c$ superconductors by
comparing the two families.  To data, it is probably the most
crucial issue to confirm the pair-symmetry of the newly discovered
superconductor, for which theoretical scientists proposed several
possible models just after the discovery of the superconductors,
among which the multi-gaped $s$-wave is generally acceptable,
including the $s$$_{\pm}$ \cite{4,5,6} and $s$$_{++}$ wave
\cite{7,8,9}. Both states represent the same hole Fermi pockets,
while have opposite signs for the electron pockets, namely, the
$s$$_{\pm}$ wave is identified as a sign-reversal $s$-wave model,
while a non-sign-reversal for the $s$$_{++}$ state. Recently results
from various experiments can hardly get a consensus for identifying
which state is the real nature of this superconductor
\cite{10,11,12}. Meanwhile, the $d$-wave model with opposite signs
for the nearest-neighbor electron pockets still remains competing
with other models, once there are nodes on the hole pockets or even
on both the electron and hole pockets \cite{6,13,14}. More recently
results suggested that different systems in the iron-pnictide family
may represent different pair-symmetry types, even that the
pair-symmetry can be quite different from material to material
\cite{12}.  The varieties of the possible scenarios arouse further
investigations, among which the impurity substitution is one of the
most promising ways to address the issue and even to uncover
competing orders, because the pair-breaking mechanism from both
magnetic and nonmagnetic impurities should be different for these
models.

The iron-pnictide superconductor contains as common Fe$_2$$X$$_2$
($X$=As, P or Se) planes, which is well-known as the superconducting
layer.  The substitution of point defect impurities for Fe is
introduced for understanding the physical properties, like what was
comprehensively studied in cuprates.  According to Anderson's
theorem, the nonmagnetic impurity (NMI) cannot break pairing in an
isotropic SC gap but for an anisotropic gap \cite{15}, while the
pair-breaking effect of the magnetic impurities is independent of
gap type.  Thus, the nonmagnetic point defects are of great
important.  The Zn$^{2+}$ with tightly closed $d$-shell is preferred
as an ideal NMI \cite{16}.  Typically, Zn substitution for Cu was
carried out over the last two decades on the cuprate superconductors
such as YBa$_2$Cu$_3$O$_{7-\delta}$ \cite{16,17,18},
(La,Sr)$_2$CuO$_4$ \cite{16,19,20,21}, and Bi$_2$Sr$_2$CaCu$_2$O$_8$
\cite{16,22,23,24}.  A few at.\% of the Zn acts as a strong
scattering center and remarkably depresses SC due to the d-wave
anisotropic gap for cuprates \cite{16,17,18,19,20,21,22,23,24}.
Since the doped Zn often plays a crucial role of pairing symmetry
determination of previous superconductors, we may expect that it
works with the Fe-based superconductor as well.

Previous Zn studies for the Fe-based superconductor seem to be
contradicted: Cheng $et~al$. reported that the doped Zn can hardly
affects SC of the $p$-type Ba$_{0.5}$K$_{0.5}$Fe$_2$As$_2$\cite{25},
as Li $et~al$. did in LaFeAsO$_{0.85}$F$_{0.15}$\cite{26}. However,
we found that the SC was completely suppressed by at most 3 at.\% of
Zn for LaFeAsO$_{0.85}$ once using high-pressure method \cite{27}.
Comparable result was obtained in the
K$_{0.8}$Fe$_{2-y-x}$Zn$_x$Se$_2$ superconductors\cite{28}. Since
the Zn substitution generally suffered from issue of the low melting
point and high volatility\cite{23,24}, it is uncertain that whether
Zn has been successfully substituted into the Fe-site for previous
polycrystals synthesized in ambient-pressure. Our recent studies
indeed showed that more than 2 at.\% of Zn were hardly doped into
the $n$-type Ba(Fe,Co)$_2$As$_2$ superconductor at ambient pressure
condition\cite{29}.  In contrast, linear $T$$_c$ suppression was
found for the high-pressure prepared
BaFe$_{2-2x-2y}$Zn$_{2x}$Co$_{2y}$As$_2$ superconductors\cite{30}.
To avoid overestimation of the net Zn we proposed growing highly
Zn-doped single crystals of the Fe-based superconductor by using a
high-pressure technique.

In this study, we studied the impurities effect on the $p$-type
(Ba,K)Fe$_2$As$_2$ superconductors by a high-pressure and
high-temperature method, for which magnetic and nonmagnetic elements
around Fe were selected as the dopant, including 3$d$ metals of Mn,
Co, Ni, Cu and Zn, and Ru from 4$d$. The specific heat, magnetic and
transport properties indicate that the SC is robustly against
impurity of Ru, while weakly against the impurities of Mn, Co, Ni,
Cu, and Zn.
\section{experimental}
Single-crystalline samples of Ba$_{0.5}$K$_{0.5}$Fe$_2$As$_2$ (BK)
and Ba$_{0.5}$K$_{0.5}$Fe$_{2-2x}$$M$$_{2x}$As$_2$ ($M$ = Mn, $x$ =
0.02 and 0.05; $M$ = Ru, Co, Ni, Cu and Zn, nominal $x$ = 0.05, 0.10
and 0.15) were prepared in a high-pressure apparatus as reported
elsewhere \cite{30}.  Here the start materials are BaAs (lab made),
KAs (lab made), FeAs (lab made), Fe (3N), Mn ($>$99.9\%), Ru
($>$99.9\%), Co ($>$99.5\%), Ni ($>$99.99\%), Cu ($>$99.9\%) and Zn
(4N).  Note that the pellet was self-separated into sizes of around
0.3$\times$0.2$\times$0.1 mm$^3$ or much smaller after it left in
vacuum for 2-3 days. The selected single crystals were held on a MgO
substrate with $ab$-plane parallel with the substrate, and then
cleaved into thin slices along $c$-axis as discussed in early report
\cite{30}.  To confirm the impurity substitution the crystals were
measured in an electron probe micro-analysis (EPMA, JXA-8500F, JEOL)
soon after cleaved. Table 1 gives the real value of $x$ for
Ba$_{0.5}$K$_{0.5}$Fe$_{2-2x}$$M$$_{2x}$As$_2$ ($M$ = Mn, Ru, Co,
Ni, Cu and Zn) with start value of $x$ = 0.05.  The result
demonstrates little difference from the starting materials, although
a slightly less concentration for Mn, Ru, Ni and Zn. However, we
will use the real concentration of $x$ for the following analysis.

The cleaved single crystals were also studied by x-ray diffraction
(XRD) method in the Rigaku Ultima-IV diffractometer using
CuK$\alpha$ radiation. The single crystals were also ground and
studied by a powder XRD method, and the results indicated that the
tetragonal ThCr$_2$Si$_2$-type structure ($I$4/$mmm$) is formed over
the compositions without second phase \cite{2,31}.

In the DC magnetic susceptibility ($\chi$) measurement, since the
size of an individual crystal is too small to obtain accurate
measurements, we loosely gathered small crystals ($\sim$30 mg in
total) into a sample holder, and measured in Magnetic Properties
Measurement System, Quantum Design. The sample was cooled down to 2
K without applying a magnetic field (zero-field-cooling, ZFC),
followed by warming to 45 K in a field of 10 Oe and then cooled down
again to 2 K (field-cooling, FC).

The cleaved single crystals were used for the in-plane DC electrical
resistivity ($\rho$$_{ab}$) measurement. To minimize the structure
defects of the single crystals, we cleaved the crystals to
$\sim$1-10 $\mu$m in thickness and cut them into a quadrate slices
shape as $\sim$100$\times$50 $\mu$m$^2$.  And then four terminals of
the cleaved $ab$-plane were pasted with platinum wires by using
silver paste.  The $\rho$$_{ab}$ was measured between 2 K and 300 K
in Physical Properties Measurement System - 9 T, Quantum Design.
Such cleaved single crystals were also measured the Hall coefficient
($R$$_H$) in PPMS, where the electric current was along the
$ab$-plane and $H$ was applied parallel to $c$-axis. For the each
sample with amount of 12-14 mg crystal, we measured the specific
heat ($C$$_p$) in PPMS-9T from 2 to 300 K by a heat-pulse relaxation
method.

\begin{table*}[ht]
\tabcolsep 0pt \caption{ The columns give the parameters (from left
to right) of Ba$_{0.5}$K$_{0.5}$Fe$_{2-2x}$$M$$_{2x}$As$_2$ ($M$ =
Fe, Mn, Ru, Co, Ni, Cu and Zn, nominal $x$=0.05): real atomic
concentration of $M$($x$) from the EPMA measurement, lattice
parameters of $a$ and $c$ from powder XRD, $T$$_{c\rho}$ from
resistivity data, and $\Delta$$C$$_p$/$T$$_{c\rho}$. The samples of
Ba$_{0.5}$K$_{0.5}$Fe$_{2-2x}$$M$$_{2x}$As$_2$ ($M$ = Fe, Mn, Ru,
Co, Ni, Cu and Zn) are abbreviated to BK, BK-Mn, BK-Ru, BK-Co,
BK-Ni, BK-Cu and BK-Zn,respectively.} \vspace*{-12pt}
\begin{center}
\def\temptablewidth{1.0\textwidth}
{\rule{\temptablewidth}{1pt}}
\begin{tabular*}{\temptablewidth}{@{\extracolsep{\fill}}ccccccc}
Samples&

 $M$($x$)

 &$a$ (\AA)&   $c$ (\AA) &

 $T$$_{c\rho}$ (K)

& $\Delta$C$_p$/$T$$_{c\rho} $(mJ mol$^{-1}$ K$^{-2}$)

 \\   \hline
     BK  & /       & 4.014(2)         & 13.298(2) & 37.78 & 44.50  \\
      BK-Mn     & 0.039(2)      &     3.984(1)   & 13.196(3) & 9.53 & /   \\
      BK-Ru    & 0.032(6)       &  4.051(1)   & 13.419(4) & 37.14 & 73.49  \\
      BK-Co& 0.052(2) & 4.038(1) &13.383(4) &30.31 &39.26  \\
       BK-Ni & 0.039(4)& 3.990(1) &13.229(1) &26.75 &/ \\
       BK-Cu& 0.044(1)& 3.970(1) &13.050(5) &22.29 &/\\
       BK-Zn& 0.040(2)& 4.102(2) &13.322(3) &30.15 &21.66\\
       \end{tabular*}
       {\rule{\temptablewidth}{1pt}}
       \end{center}
       \end{table*}
\section{results and discussion}
\subsection{X-ray diffraction}
The XRD patterns for the cleavage plane of the separated crystals
Ba$_{0.5}$K$_{0.5}$Fe$_{1.9}$$M$$_{0.1}$As$_2$ ($M$ = Fe, Mn, Ru,
Co, Ni, Cu and Zn, which are abbreviated as BK, BK-Mn, BK-Ru, BK-Co,
BK-Ni, BK-Cu and BK-Zn, respectively) are shown in Fig. 1(a).  The
obvious orientation toward [0 0 2$n$] ($n$ is integer) indicates
that the cleavage plane is the $ab$-plane of the ThCr$_2$Si$_2$-type
structure. Note that the main peak (0 0 8) for every impurity-doped
crystals indicates an obvious shift in 2$\theta$ as shown in Fig.
1(b), suggesting that the impurities were indeed doped into the
crystal lattice.  The lattice parameters obtained by assuming the
same structure for the powder XRD data are summarized in Table 1, as
can be seen that the impurity-doping results in change for lattice
parameters of both $a$ and $c$.  The unsystematic change in peak
shift and lattice parameters seem unlike due to the basic change in
size of doping ions. However, the difference between Fe-As and
$M$-As bond size was considered as a crucial factor as discussed in
Ref. 15. In addition, a magnetic effect is possibly included in the
$c$-axis expansion \cite{32}, especially for the nonmagnetic Zn
ions, which results in nearly isotropic expansion for both $a$ and
$c$. Comparably, Zn-doped
BaFe$_{1.91-x}$Zn$_x$Co$_{0.11}$As$_2$\cite{30} and
YBa$_2$Cu$_{3-3x}$Zn$_{3x}$O$_{7-\delta}$ \cite{33} also results in
an isotropic expansion of the lattice.
\begin{figure}
\includegraphics*[bb=0 0 1275 555,width=0.48\textwidth]{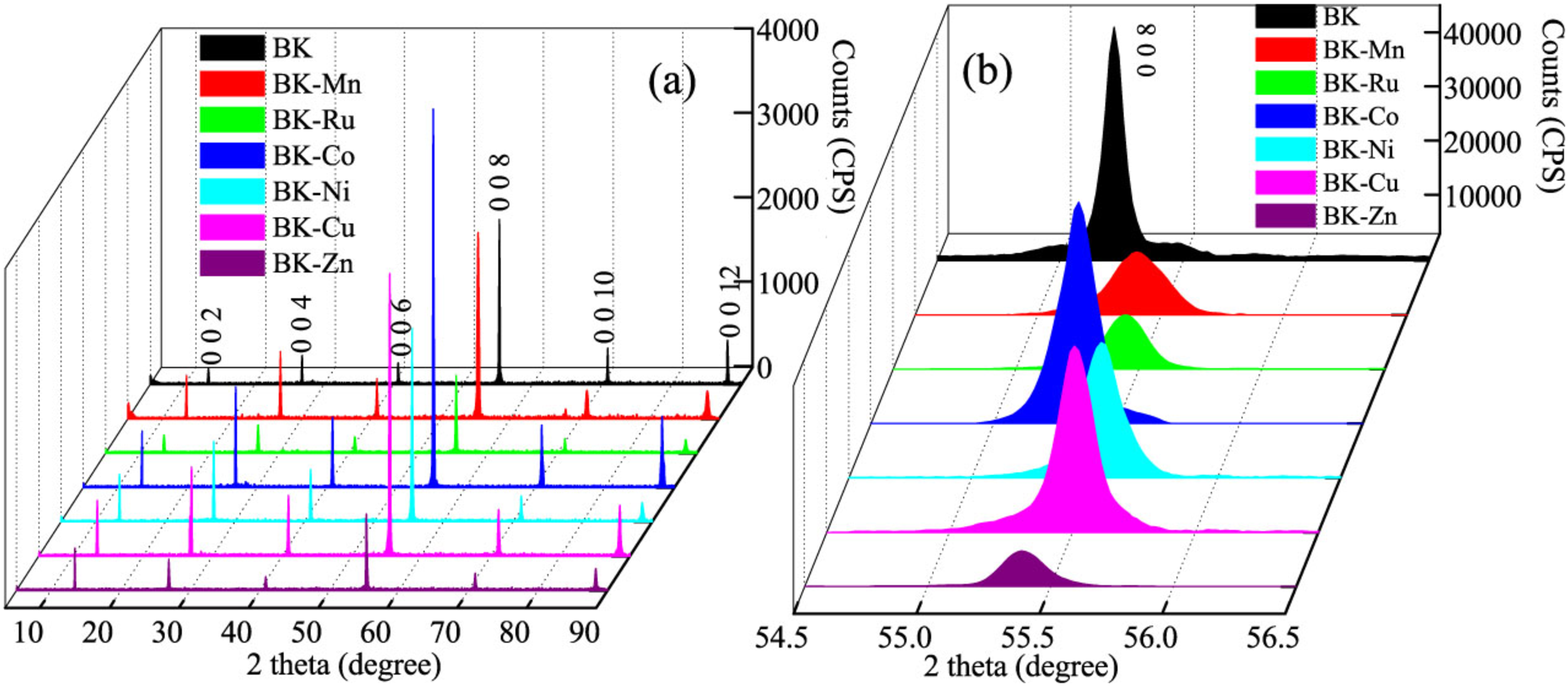}
\caption{XRD pattern of the single crystals
Ba$_{0.5}$K$_{0.5}$Fe$_{2-2x}$$M$$_{2x}$As$_2$ ($M$= Mn, Ru, Co, Ni,
Cu and Zn, real $x$ values are shown in Table 1).} \label{fig1}
\end{figure}
\subsection{Magnetic measurement}
Fig. 2 shows $T$ dependence of the Ba$_{0.5}$K$_{0.5}$Fe$_2$As$_2$
and Ba$_{0.5}$K$_{0.5}$Fe$_{2-2x}$$M$$_{2x}$As$_2$ ($M$ = Mn, Ru,
Co, Ni, Cu and Zn), where the impurity concentration of $x$ is
obtained from the EPMA measurements. The host crystal BK shows the
maximum $T$$_c$ of 38 K as reported elsewhere \cite{2}. Obviously,
the SC of Ba$_{0.5}$K$_{0.5}$Fe$_2$As$_2$ shows strong against with
the Ru impurity, which is accordance with the previous studies on
the Ru substitution effect on LaFeAsO$_{1-x}$F$_x$ \cite{34} and
NdFeAsO$_{0.89}$F$_{0.11}$ superconductors \cite{35}.  The magnetic
impurity of Mn indicates the sharpest $T$$_c$ suppression among all
impurities. It is surprising that the $T$$_c$-reduction effect from
the 3$d$ transition metals (Co, Ni, Cu and Zn) are similar each
other, regardless of magnetic or nonmagnetic impurities.
\begin{figure}
\includegraphics*[bb=0 0 1230 1311,width=0.48\textwidth]{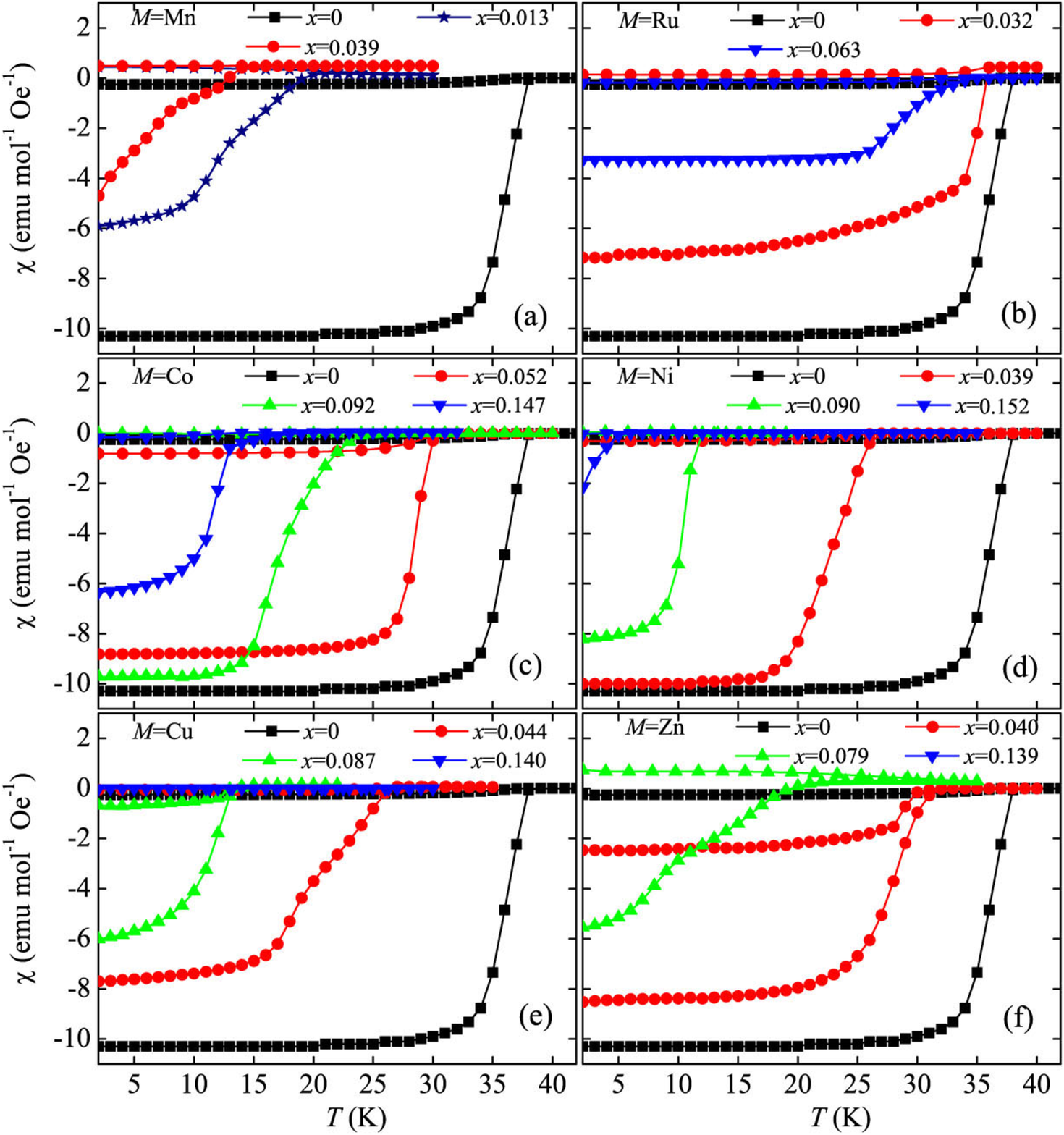}
\caption{ $\chi$ vs. $T$ for
Ba$_{0.5}$K$_{0.5}$Fe$_{2-2x}$$M$$_{2x}$As$_2$  ($M$ = Mn, Ru, Co,
Ni, Cu and Zn) at $H$ = 10 Oe.} \label{fig2}
\end{figure}
\subsection{Transport property}
\begin{figure}
\includegraphics*[bb=0 0 1270 1346,width=0.48\textwidth]{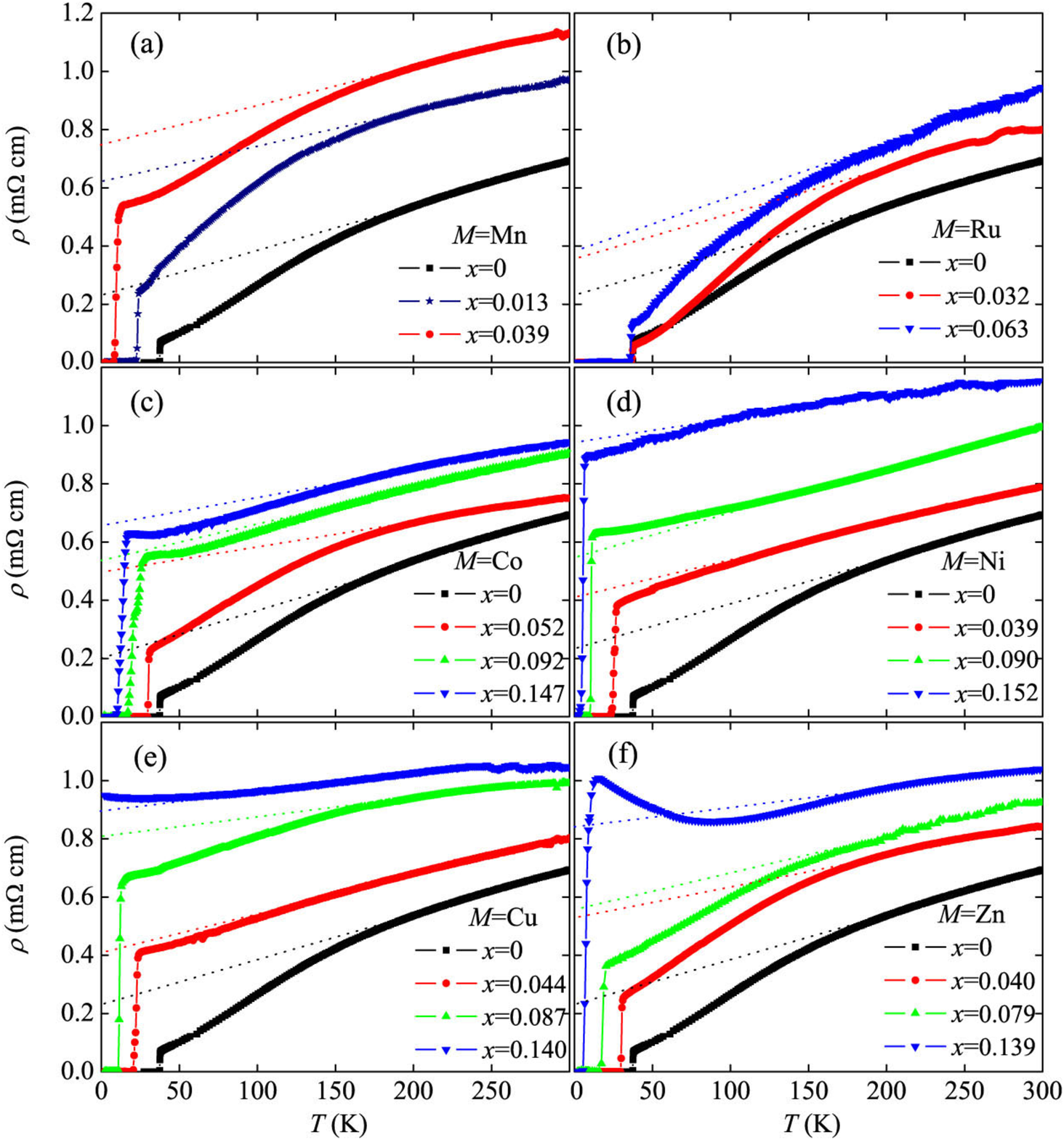}
\caption{$\rho$   vs. $T$ for
Ba$_{0.5}$K$_{0.5}$Fe$_{2-2x}$$M$$_{2x}$As$_2$ ($M$ = Mn, Ru, Co,
Ni, Cu and Zn).} \label{fig3}
\end{figure}
Transport properties provide direct information for the influence of
impurities or defects on various SC properties, including the
carrier properties, coupling between charges, spin degrees of
freedom, and more importantly the pair-breaking symmetry
\cite{16,17}. To obtain a reliable transport data high-quality
single crystals are essential with substitution of impurities.  Fig.
3 shows the temperature dependence of $ab$-plane resistivity
($\rho$) for the Ba$_{0.5}$K$_{0.5}$Fe$_{2-2x}$$M$$_{2x}$As$_2$ ($M$
= Fe, Mn, Ru, Co, Ni, Cu and Zn). The $T$$_c$ was defined from the
peak value for the plots of $d\rho$ /$dT$ vs. $T$.  It is clearly
observed that $T$$_c$ goes down with substitution of Mn, Co, Ni, Cu
and Zn, while is weekly suppressed by Ru, well accordance with the
magnetic results.  Note that for the
Ba$_{0.5}$K$_{0.5}$Fe$_{2-2x}$Mn$_{2x}$As$_2$ ($x$ = 0, 0.013 and
0.039), the  $\rho$-$T$ curves are almost parallel each other at the
high-$T$ region as above $T$$_c$.  Such behavior establishes that
the hole content is modified by the defects rather than the electron
irradiation.  At low-$T$ on the other hand, an upturn in the
$\rho$-$T$ curve is observed with substitution of defect content
($x$ $<$ 0.05). This phenomenon has been often interpreted as the
occurrence of localization.  In case of Ru-doped crystals, the
$\rho$-$T$ curves show almost parallel upturn with substitution of
Ru at both high- and low-$T$ regions, suggesting the absence of
localization. The $\rho$-$T$ curves for the Co, Ni, Cu and Zn-doped
crystals are observed no parallel shift from that of the
impurity-free crystal.  However, the low-$T$ upturns of the
resistivity appear for the impurity-doped crystals due to
localization, regardless of magnetic or nonmagnetic impurities.
Typically, the high-level Zn-doped crystals
Ba$_{0.5}$K$_{0.5}$Fe$_{2-2x}$Zn$_{2x}$As$_2$ ($x$ = 0.139)
indicates a dramatically low-$T$ upturn from localization, similar
phenomenon was observed in the Zn-substituted
$A$(Fe,Zn,Co)$_2$As$_2$ superconductors \cite{30,36}.

As the resistance of the superconductor shows a metal-like behavior,
it decreases linearly with temperature at
 high-temperature regions. Therefore, we define the residual resistivity
 $\rho_0$ by the extrapolation of $T$-linear resistivity to 0 K for the
 linear $T$-dependence at high-$T$ region.  The residual resistivity $\rho_0$ gradually increased
 with increasing doping level except Ru, and the increasing rate of $\rho_0$ with $x$
 are $\sim$98.2, 22.3, 42.8, 46.2 and 35.1  $\mu\Omega$cm/$\%$ for Mn, Co, Ni, Cu and Zn, respectively.
 The residual resistivity is due to defect scattering, although it is not
 easy to obtain accurate determinations of the scattering rate directly from
 resistivity data, an alternative approach is to seek information from the decrease
 of $T$$_c$ induced by the scattering centers \cite{16}.  Fig. 4 shows the residual
 resistivity $\rho_0$ dependence of $\Delta$$T$$_c$, where the $T$$_c$ data are from resistivity measurements.
 The $T$$_c$ is gradually suppressed with increasing $\rho_0$  for  all impurities except Ru.
 The $T$$_c$ is nearly independent of $\rho_0$ for the substitution of Ru,
 while suppressed by impurities of Mn, Co, Ni, Cu and Zn as suppression rate of 66.77, 76.78, 46.43, 51.67 and 59.45 K/m$\Omega$cm, respectively.
 Note that these impurities are observed as similar suppression rate.
 However, the theoretical residual resistivity per 1\% impurity with
 delta-functional strong potential is just $\sim$20 $\mu\Omega$cm,
 and SC will also vanish with doping 1\% of either magnetic or
 nonmagnetic impurities for the $s$$_{\pm}$ wave model \cite{7,8,9}.
 Consequently, the suppression rate is around 1900 K/m$\Omega$cm,
 indicating that the impurity scattering cross section is enlarged by the many-body effect,
  other than the pair-breaking effect \cite{37}, which we will discuss in detail in the discussion part.

\begin{figure}
\includegraphics*[bb=0 0 673 494,width=0.48\textwidth]{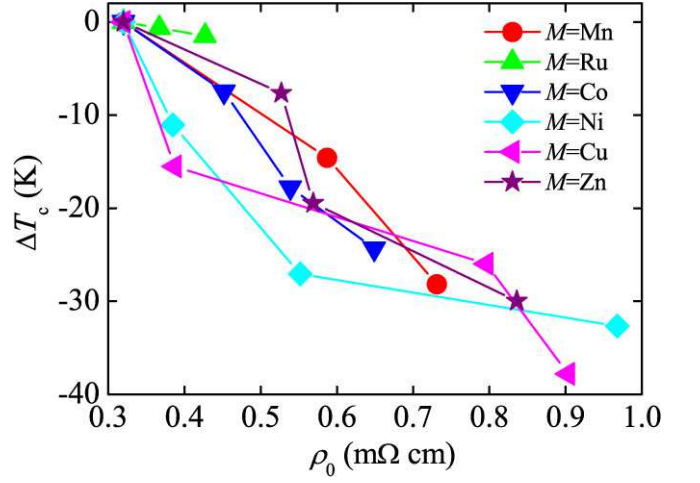}
\caption{$\Delta$$T$$_c$ as a function of residual resistivity
($\rho$$_0$) for the superconductors of
Ba$_{0.5}$K$_{0.5}$Fe$_{2-2x}$$M$$_{2x}$As$_2$ ($M$= Mn, Ru, Co, Ni,
Cu and Zn).} \label{fig4}
\end{figure}

\begin{figure}
\includegraphics*[bb=0 0 1346 478,width=0.48\textwidth]{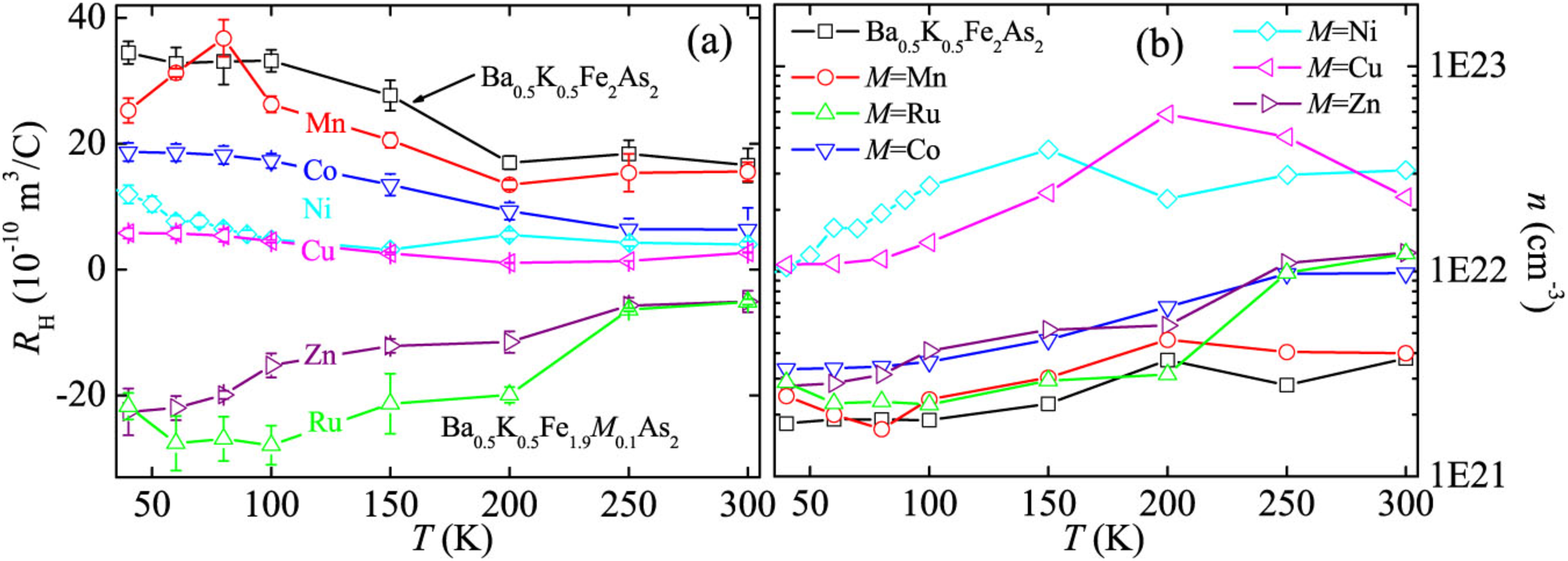}
\caption{ (a) Hall coefficient ($R$$_H$) vs. $T$ and (b) carrier
density ($n$) vs. $T$ for single-crystalline
Ba$_{0.5}$K$_{0.5}$Fe$_{2-2x}$$M$$_{2x}$As$_2$ ($M$=Fe, Mn, Ru, Co,
Ni, Cu and Zn, real $x$ values are shown in Table 1). } \label{fig5}
\end{figure}

\begin{figure*}
\includegraphics*[bb=0 0 1440 760,width=1.0\textwidth]{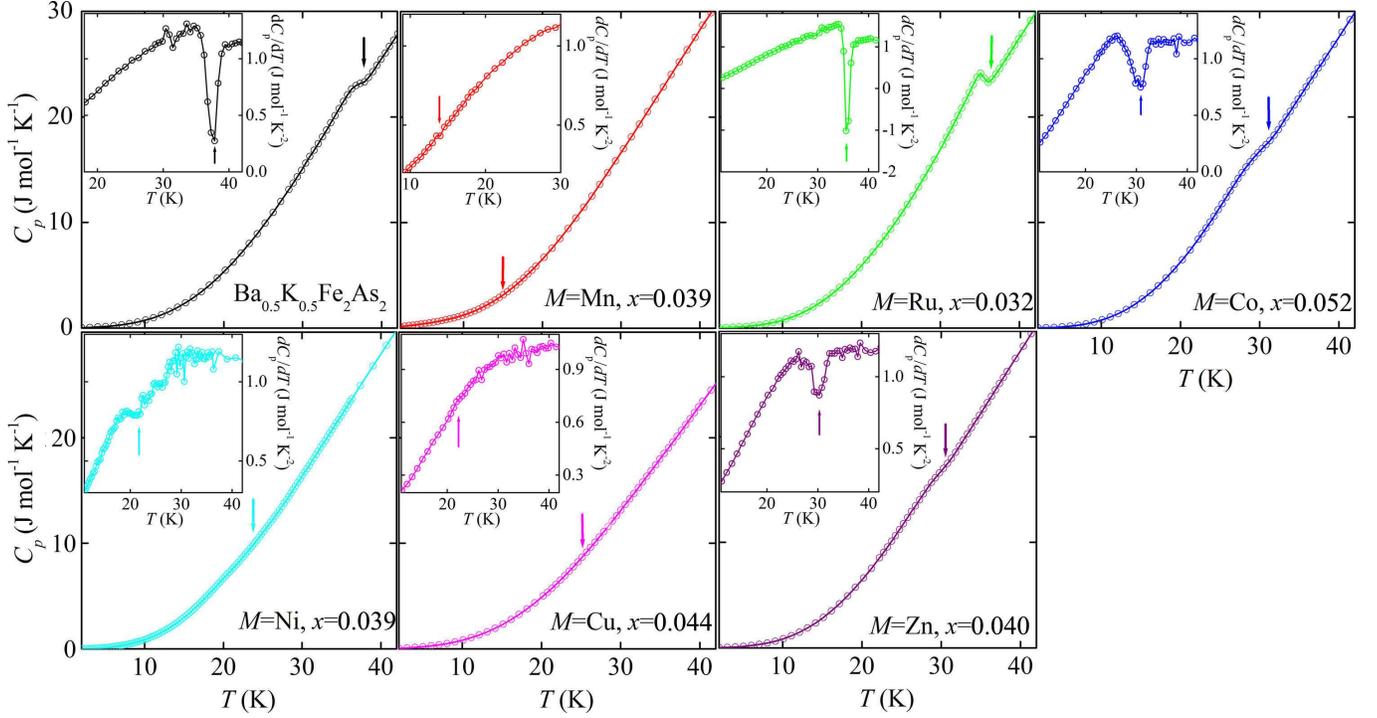}
\caption{ Specific heat dependence of the temperature for
Ba$_{0.5}$K$_{0.5}$Fe$_{2-2x}$$M$$_{2x}$As$_2$ ($M$ = Mn, Ru, Co,
Ni, Cu and Zn), where inset of each figure demonstrates the
derivation of $C$$_p$ to $T$, and the arrows indicate the heat
capacity anomaly. } \label{fig6}
\end{figure*}

\begin{figure*}
\includegraphics*[bb=0 0 1440 735,width=1.0\textwidth]{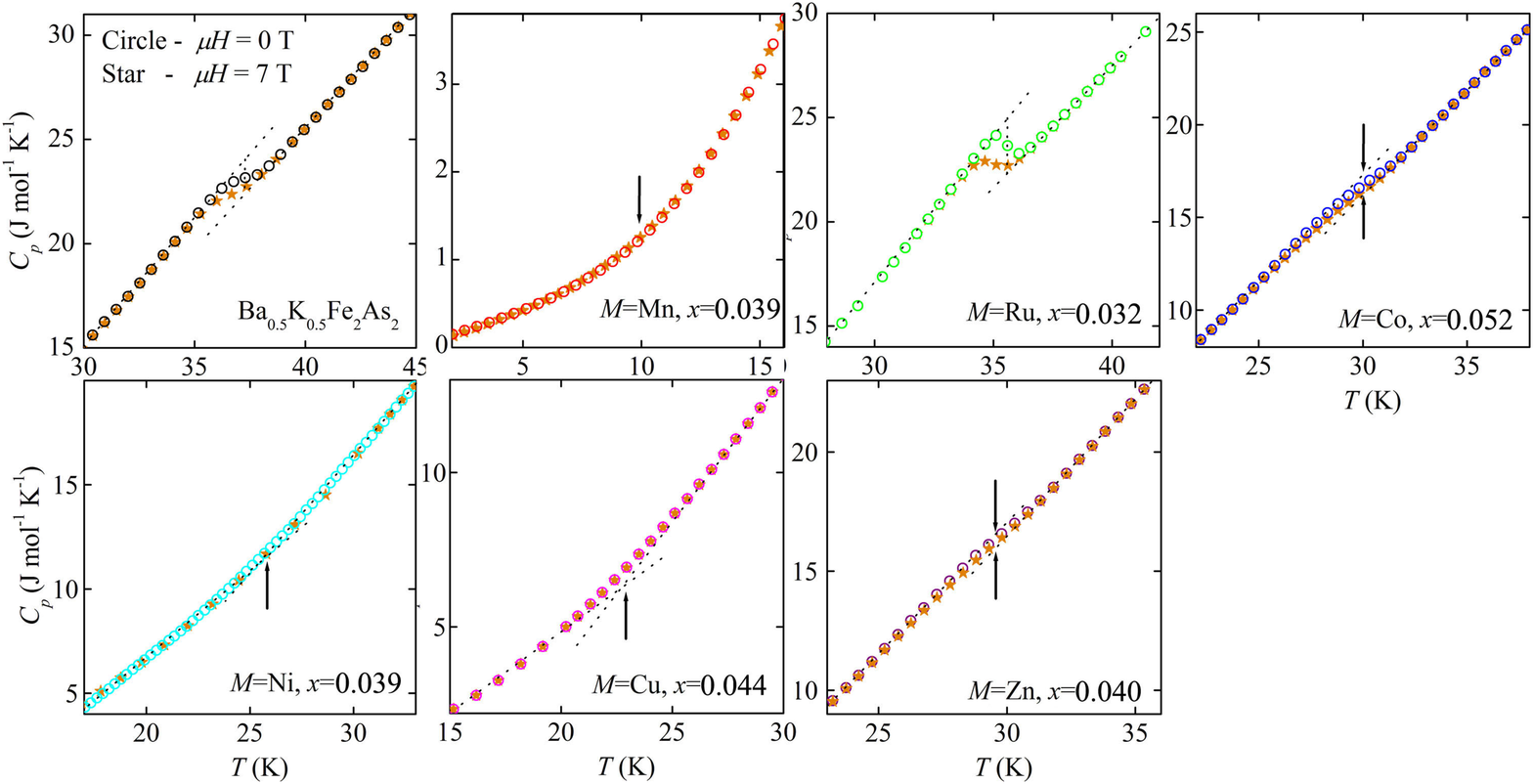}
\caption{Specific heat dependence of the temperature for
Ba$_{0.5}$K$_{0.5}$Fe$_{2-2x}$$M$$_{2x}$As$_2$ ($M$ = Mn, Ru, Co,
Ni, Cu and Zn) with and without magnetic field of 7 T. }
\label{fig7}
\end{figure*}
Fig. 5 shows the $T$ dependence of Hall coefficient ($R_H$) and
carrier density ($n$) for the BK, BK-Mn, BK-Ru, BK-Co, BK-Ni, BK-Cu
and BK-Zn single crystals. The data for the impurity-free crystal
accesses to the early data [25,37]. With substitution of 5 at.\% of
Mn, Co, Ni or Cu, the BK crystal is observed slightly reducing in
$R_H$, while increasing carrier density. Sato and co-workers
\cite{10,34,35} proposed that the decrease in the absolute magnitude
of $R$$_H$ is due to the weakening of the magnetic fluctuations, as
in the case of the thermoelectric power $S$. However, it is
surprisingly that the impurities of Ru and Zn result in a negative
$R$$_H$, which seems like that the introduction of Ru and Zn ions
yield the charge carrier type from hole-doping to electron-doping.
For the normal state, we found there is no significant change over
various substitutes, indicating that the transition-metal
substitution do not substantially alter the actual carrier density.
This is reasonable because the substitution is isovalent.  Regarding
the previous impurity effect on charge carrier of both Fe-based and
Cu-based superconductors, fairly little change was observed in the
$R$$_H$ measurements as well \cite{30,33}. The actual carrier
density change by transition-metal impurities does not account for
the systematic $T$$_c$ decrease \cite{5}.
\subsection{Specific heat data}
The temperature dependent specific heat ($C_p$) in zero-field for
the BK, BK-Mn, BK-Ru, BK-Co, BK-Ni, BK-Cu and BK-Zn are given in
Fig. 6, where inset of each figure demonstrates the derivation of
$C_p$ to $T$. Obvious heat capacity anomaly, indicated by the arrow,
is associated well with the SC transition temperature for BK, BK-Ru,
BK-Co, BK-Ni, and BK-Zn. However, there is almost absence of anomaly
at $T$$_c$ for the BK-Mn and BK-Cu.
\begin{figure}
\includegraphics*[bb=0 0 653 487,width=0.48\textwidth]{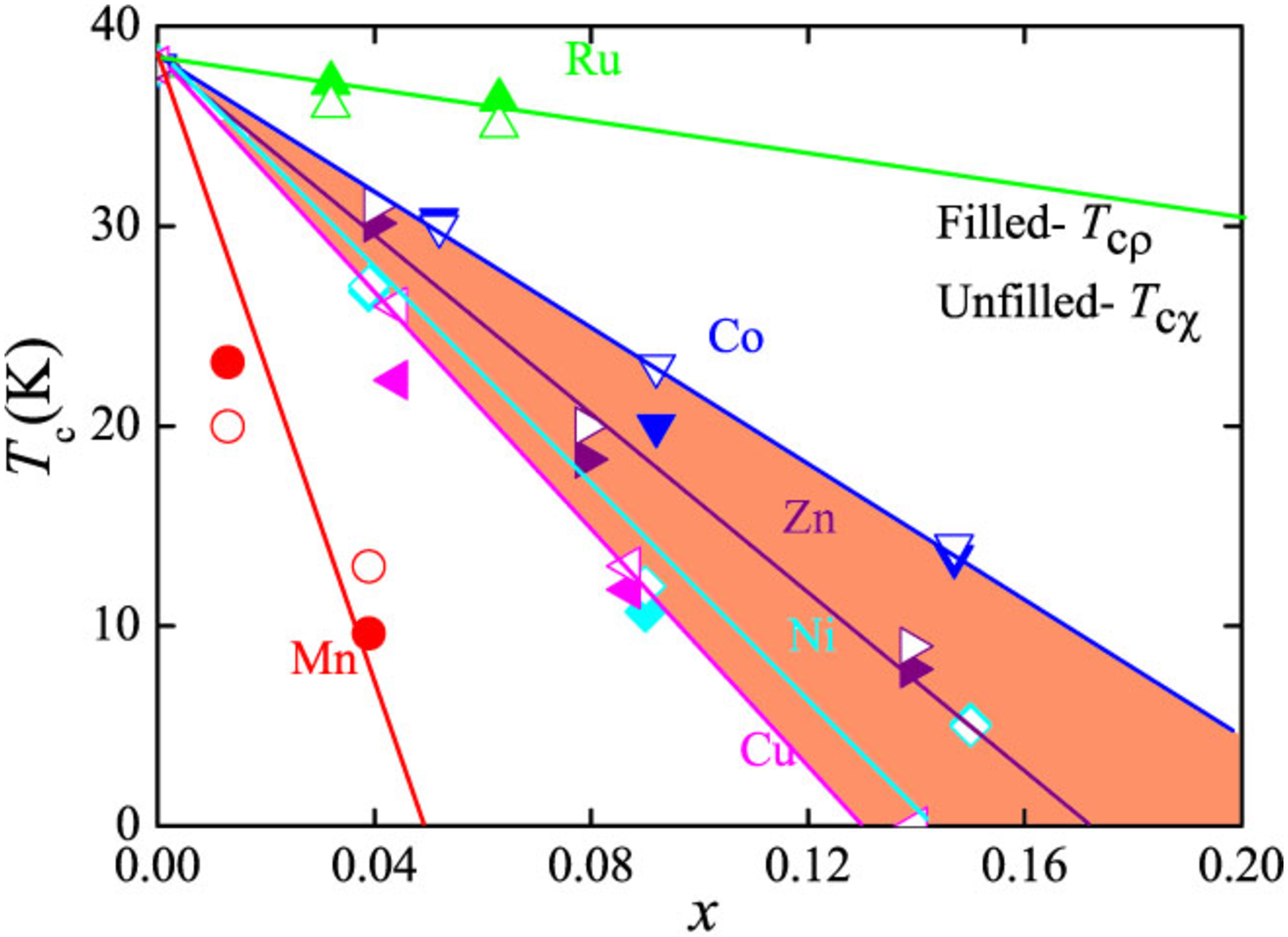}
\caption{$T$$_c$ vs. $x$ for the superconductors of
Ba$_{0.5}$K$_{0.5}$Fe$_{2-2x}$$M$$_{2x}$As$_2$ ($M$= Mn, Ru, Co, Ni,
Cu and Zn).} \label{fig8}
\end{figure}
It is possible that disorder regarding impurity distribution causes
inhomogeneous SC states, much broadening the expected peak, and the
broad anomaly is masked by the lattice contributions \cite{39,40}.
In addition, it was found that the character of the anomaly is
strongly doping dependence \cite{41}. However, the reason for the
absence of anomaly in Mn-, Ni-, and Cu-substituted samples need
further investigation. Fig. 7 shows the $C$$_p$-$T$ curves in both
fields of 0 and 7 T at around $T$$_c$, from which we estimate the
specific heat jump ($\Delta$$C_p$/$T$$_{c\rho}$) for these
transitions at zero-field as shown in Table 1, where $T_{c\rho}$ is
the $T$$_c$ estimated from resistivity data. It is observed that the
impurities substitution change weekly on superconducting phase as
judged from the size of specific heat jump, although the Co and Zn
substitution reduced weekly on $\Delta$$C_p$/$T_{c\rho}$, as well as
the Ru-doping enhances the $\Delta$$C_p$/$T_{c\rho}$ (73.49 mJ
mol$^{-1}$ K$^{-2}$) to about two times of the impurity-free sample
(44.50 mJ mol$^{-1}$ K$^{-2}$). On the other hand, the applied
magnetic field of 7 T is not large enough to suppress the anomaly
(see Fig. 7) due to the high upper critical fields ($>$ 55 T). Since
both the superconducting temperature and the upper critical fields
in these superconductors are relatively high, we can hardly make a
reliable estimate of the normal state electronic specific heat.
\begin{figure}
\includegraphics[bb=0 0 687 483,width=0.48\textwidth]{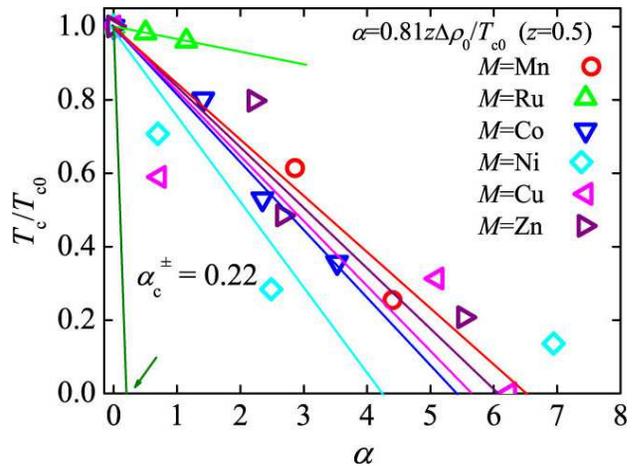}
\caption{ $T$$_c$/$T$$_{c0}$ vs. $\alpha$  with various calculations
for Ba$_{0.5}$K$_{0.5}$Fe$_{2-2x}$$M$$_{2x}$As$_2$ ($M$= Mn, Co, Ni,
Cu and Zn). The $T$$_c$ of each impurity-doped sample is normalized
with $T$$_{c0}$ of the impurity-free compound.  The pair-breaking
rate $\alpha$ is estimated as $\alpha$ =
0.88$z$$\Delta$$\rho$$_0$/$T$$_{c0}$, where $\Delta$$\rho_0$ is the
difference of the residual resistivity from that of impurity-free
crystals, and $z$ is the renormalization factor, for which we take
$z$ = 0.5 from ARPES in 122 superconductor\cite{43,44}.}
\label{fig9}
\end{figure}
\section{discussion}
We have described the influence of impurities on the magnetic,
transport and specific heat properties in the
Ba$_{0.5}$K$_{0.5}$Fe$_2$As$_2$ superconductor.  On basis of these
results, we focus on the discussion of pair-breaking effects in
terms of both $s$$_{\pm}$ and $s$$_{++}$ wave states.

Based on density functional calculations it was found that the
impurity effects in iron-based superconductors can be classified
into three groups according to the derived parameters: (i) Mn (0.3
eV), Co (-0.3 eV), and Ni (-0.8 eV), (ii) Ru (0.1 eV) and (iii) Zn
(-8 eV) \cite{42}. Among these impurities the nonmagnetic Zn works
as a unitary scattering potential that is comparable to the
bandwidth, as a result of the quite strong potential. Consequently,
it is expected as strictly pair-breaking effect on the anisotropic
SC gaps. According to Ref. 7 the reduction in $T$$_c$ due to strong
impurity potential in the $s$$_{\pm}$ wave state is $\sim$50$z$
K/\%, where $z$ is the renormalization factor (= $m$/$m$$^*$; $m$
and $m$$^*$ are the band-mass and the effective mass, respectively).
The effective mass was estimated as 2$m$$_e$ from ARPES in 122
superconductor \cite{43,44}; thus we obtain the suppression rate of
25 K/\% for z = 0.5.  In contrast, the $T$$_c$ would be weakly
suppressed by impurities in the $s$$_{++}$ wave state, due to the
following reasons: (i) suppression of the orbital fluctuations,
which is a possible origin of the $s$$_{++}$ wave state, because of
the violation of the orbital degeneracy near the impurities, and
(ii) the strong localization effect in which the mean-free-path is
comparable to the lattice spacing \cite{7}.  These may account for
the observed $T$$_c$ decrease.  In our present results, the decrease
of the $\chi$  and $\rho$-defined $T$$_c$ ($T$$_{c\chi}$  and
$T$$_{c\rho}$ ) with doping level $x$ for the superconductors of
Ba$_{0.5}$K$_{0.5}$Fe$_{2-2x}$$M$$_{2x}$As$_2$ ($M$= Mn, Ru, Co, Ni,
Cu and Zn) is given in Fig. 8. We estimated the suppression rate of
Zn is 2.22 K/\% by applying a linear function to the $T$$_{c\rho}$
vs. $x$, which is well accordance with the
BaFe$_{1.89-2x}$Zn$_{2x}$Co$_{0.11}$As$_2$ superconductors\cite{30}.
The observed robustness of SC seems like to contradict with the
nonmagnetic impurity quantitatively in the $s$$_{\pm}$ wave model.
Applied a linear function to the $T$$_c\rho$ -$x$, the suppression
rates for Mn, Ru, Co, Ni, and Cu are 6.98, 0.27, 1.73, 2.21 and 2.68
K/\%, respectively. Among these impurities Mn is observed as the
strongest suppression effects, even though such influence is quite
weaker than what was expected from the $s$$_{\pm}$ wave model.  The
negligent suppression effect from Ru in present compound is
consistent with in the 1111-system \cite{10,45}. The other
transition metal impurities show less difference in suppression
effect with Zn.

On basis of previous pair-breaking analysis in the
BaFe$_{1.89-2x}$Zn$_{2x}$Co$_{0.11}$As$_2$ superconductors
\cite{30}, we calculated the pair-breaking rate $\alpha$ =
$z\hbar\gamma$/2$\pi$k$_B$$T$$_{c0}$ for
Ba$_{0.5}$K$_{0.5}$Fe$_{2-2x}$$M$$_{2x}$As$_2$ ($M$ = Mn, Ru, Co,
Ni, Cu and Zn), where $T$$_{c0}$ is the $T$$_c$ of the impurity-free
compound, and $\gamma$ is the electron scattering rate. Previous
theoretical study proposed a relation between $\gamma$ and
$\Delta\rho_0$ as $\Delta\rho_0$ ($\mu\Omega$cm) = 0.18$\gamma$ (K)
in terms of five-orbital model for 122 systems, here $\Delta\rho_0$
is the difference of the residual resistivity of the impurity-doped
and impurity-free crystals.  For the $s$$_{\pm}$-wave state, the SC
should vanish in the range $\alpha$ $>$ $\alpha^{\pm}_{c}$= 0.22
\cite{7}. For the present experiment, we estimated $\alpha$ =
0.88z$\Delta\rho_0$/$T$$_{c0}$ by using $z$ = 0.50 as shown in Fig.
9. The $T$$_c$/$T$$_{c0}$ vs. $\alpha$ data change in roughly
linear; thereby we applied a linear function to the data and
estimated the critical pair-breaking parameters as 6.52, 5.23, 4.24,
5.41 and 6.05 for impurities of Mn, Co, Ni, Cu and Zn, respectively.
Comparably result was obtained for the pair-breaking effect of Zn in
the BaFe$_{1.89-2x}$Zn$_{2x}$Co$_{0.11}$As$_2$ superconductors as
$\alpha$ = 11.49 with $z$ = 0.5. Resent data for the proton
irradiated Ba(Fe,Co)$_2$As$_2$ showing similar results as those of
our chemical doping \cite{46}. Obviously, the pair-breaking
parameters experimentally estimated for the present compound are far
above the limit $\alpha^{\pm}_c$ = 0.22 for the $s$$_{\pm}$-wave
model, suggesting that the realization of the $s$$_{++}$ wave state
rather than the $s$$_{\pm}$-wave model in the 122-type Fe-based
superconductor.

\vspace{6pt}
\section{conclusions}

To summarize, we have studied the superconductivity suppression
effect on Ba$_{0.5}$K$_{0.5}$Fe$_{2-2x}$$M$$_{2x}$As$_2$ single
crystals by substitution of transition-metal ($M$ = Mn, Ru, Co, Ni,
Cu, and Zn).  The superconductivity of the $p$-type iron-based
superconductor shows robustly against impurity of Ru, while weakly
against the impurities of Mn, Co, Ni, Cu, and Zn, whose $T$$_c$
suppression rate are 6.98, 1.73, 2.21, 2.68, and 2.22 K/\%,
respectively.  Mn is observed as the strongest suppression effects,
while the other transition metal impurities of Co, Ni, Cu, and Zn
show similar suppression effect regardless of magnetic or
nonmagnetic property. However, the present $T$$_c$ suppression rate
of both magnetic and nonmagnetic impurities remains much lower than
what is expected for the $s$$_{\pm}$-wave model. The temperature
dependence of resistivity data was observed an obviously low-$T$
upturn for the high-level impurity-doped crystals, which is due to
the occurrence of localization.  The relatively weak $T$$_c$
suppression effect from Mn, Co, Ni, Cu, and Zn are considered as a
result of localization rather than pair-breaking effect in
$s$$_{\pm}$-wave model. However, another scenario toward the
non-sign reversal $s$-wave model ($s$$_{++}$-wave) is more likely
for the present superconductors.

\section{acknowledgements}
We thank Dr. M. Miyakawa for high-pressure experiment and Dr. K.
Kosuda for EPMA experiment, and also thank Drs. D. Johrendt, H. B.
Wang, B. Y. Zhu, M. Sato and P. J. Pereira for valuable discussions.
This research was supported in part by the World Premier
International Research Center from MEXT, Grants-in-Aid for
Scientific Research (22246083) from JSPS, and the Funding Program
for World-Leading Innovative R$\&$D on Science and Technology (FIRST
Program) from JSPS.

\end{document}